\documentclass[preprint, superscriptaddress,showpacs,preprintnumbers,amsmath,amssymb,pra]{revtex4}
\usepackage{graphicx}
\usepackage{amsmath}\usepackage{slashed}

\begin{document}

\thispagestyle{empty}

\title{Characteristic properties of the Casimir free energy for metal films
deposited on metallic plates}

\author{
G.~L.~Klimchitskaya}

\author{
V.~M.~Mostepanenko}
\affiliation{Central Astronomical Observatory at Pulkovo of the Russian Academy of Sciences,
Saint Petersburg,
196140, Russia}
\affiliation{Institute of Physics, Nanotechnology and
Telecommunications, Peter the Great Saint Petersburg
Polytechnic University, St.Petersburg, 195251, Russia}

\begin{abstract}
The Casimir free energy and pressure of thin metal films deposited on metallic plates are
considered using the Lifshitz theory and the Drude and plasma model approaches to the role
of conduction electrons. The bound electrons are taken into account by using the complete
optical data of film and plate metals. It is shown that for films of several tens of nanometers
thickness the Casimir free energy and pressure calculated using these approaches differ by
hundreds and thousands percent and can be easily discriminated experimentally.
According to our results, the free energy of a metal film does not vanish in the limiting case
of ideal metal if the Drude model approach is used in contradiction with the fact that the
fluctuating field cannot penetrate in its interior. Numerical computations of
the Casimir free energy and pressure of Ag and Au films deposited on Cu and Al plates have
been performed using both theoretical approaches. It is shown that the free energy of a
film can be both negative and positive depending on the metals used. For a Au film on a Ag
plate and vice versa the Casimir energy of a film changes its sign with increasing film
thickness. Applications of the obtained results for resolving the Casimir puzzle and the problem
of stability of thin films are discussed.
\end{abstract}

\maketitle

\section{Introduction}

It is well known that the van der Waals and Casimir energies and forces arise between
two closely spaced material bodies due to the zero-point and thermal fluctuations of
the electromagnetic field \cite{1,2}. During the last few years the Casimir effect has
found numerous applications in atomic physics \cite{3,4,5,6,7}, condensed matter
physics \cite{8,9,10,11,12} and nanotechnology \cite{13,14,15} (see also reviews
\cite{16,17}). The high-tech laboratory setups using an atomic force microscope \cite{16}
and a micromechanical torsional oscillator \cite{17} made possible measuring the Casimir
interaction with unprecedented precision. This has opened up opportunities for
quantitative comparison of the experimental data with theoretical predictions of the
Lifshitz theory \cite{1,2,18}, which provides a unified fundamental description of both
the van der Waals and Casimir forces on the basis of first principles of quantum
electrodynamics and quantum statistical physics.

The Lifshitz theory represents the van der Waals and Casimir free energies and forces
as some functionals of the frequency-dependent dielectric permittivities and magnetic
permeabilities of interacting bodies. In so doing, the dielectric permittivities in
common use include the contributions from both free and bound charge carriers
(electrons for metals).    A comparison between the experimental data of all precise
experiments on measuring the Casimir force performed for metallic test bodies
\cite{19,20,21,22,23,24,25} with the predictions of the Lifshitz theory has revealed
a problem, the so-called Casimir puzzle. It turned out that if the relaxation properties
of free (conduction) electrons are taken into account in calculations (this is usually
done by using the Drude model), the theoretical results are excluded by the measurement
data at up to 99\% confidence level \cite{19,20,21,22,23,24,25}.
By contrast, if the relaxation properties of free electrons are omitted (i.e., the free
charge carriers are described by the nondissipative plasma model), the measurement
data are found in agreement with theoretical predictions at more than 90\% confidence
level \cite{26}. These results are puzzling because the major difference in theoretical
predictions of both models comes from the zero-frequency term of the Lifshitz formula,
where, according to common views, the role of dissipation is rather large and should
be taken into account.

The above puzzle gave rise to a long-term discussion in the literature, where a lot of
arguments pro and contra of using the Drude and plasma dielectric permittivities in the
Lifshitz theory has been proposed (see, e.g., Refs.~\cite{27,28,29,30,31,32,33,34}).
It should be noted that at separation distances below $1\,\mu$m, where the magnitude
of the Casimir force is large enough to be measured with high precision, the difference
in theoretical predictions using the Drude and plasma models does not exceed a few percent.
In spite of the fact that the experimental precision was equal to a fraction of a percent,
this gave grounds for attempted explanations of a puzzle by the role of some unaccounted
background effects, e.g., by the influence of patch potentials \cite{35} (thereafter the
role of this effect in the test bodies of Refs.~\cite{19,20,21,22} was shown to be
negligibly small \cite{36}).

An experimental situation has become conclusively established after the difference force
measurement was performed \cite{37,37a}, where theoretical predictions of the Drude and plasma
models differ by a factor of a few hundreds. The results of this experiment, which was
proposed in Refs.~\cite{38,39,40}, excluded the predictions of the Drude model and are
consistent with the plasma model \cite{37}. Several experiments using dielectric and
dielectric-type semiconductor test bodies \cite{6,41,43,44} are also worthy of notice.
The measurement data of these experiments exclude the predictions of the Lifshitz theory
obtained with taken into account contribution of free charge carriers (dc conductivity),
and are consistent with the same theory if the free charge carriers are omitted
\cite{6,41,43,44,45}. It is significant that the Lifshitz theory violates the third law
of thermodynamics (the Nernst heat theorem) if it describes conduction electrons in
metals with perfect crystal lattices by means of the Drude model \cite{46,47,48,49} or
takes into account the free charge carriers in dielectrics \cite{50,51,52,53}.
By contrast, if the plasma model is used for metals  and the free charge carriers are
omitted for dielectrics, the Lifshitz theory is found in perfect agreement with the
Nernst heat theorem \cite{46,47,48,49,50,51,52,53}.

As was shown recently in Ref.~\cite{54}, the Lifshitz theory using the Drude and plasma
models predicts significantly different values of the Casimir free energy and pressure
for a metallic film either sandwiched between two dielectric plates or in vacuum.
These configurations have some advantages, as compared with difference force
measurements \cite{37,38,39,40}, because in the case of the plasma model the latter
result in a detectable but rather small signal of the order of 0.1\,pN.
It was found also that when the Drude model is used the Casimir free energy does not
vanish in the limiting case of ideal metal in contradiction with the fact that
electromagnetic oscillations cannot penetrate in its interior.

In this paper, we investigate the Casimir free energy and pressure for metal films
deposited on metallic plates using the Lifshitz theory and describing the conduction
electrons by both the Drude and plasma models. The interest to this subject is twofold.
On the one hand, the configuration of two dissimilar metallic layers is analogous to
that considered in Ref.~\cite{54}. Thus, one may expect significantly different values
of the Casimir free energy and pressure when the Drude and plasma models are employed
in calculations for experimentally used film thicknesses. On the other hand, the obtained
results are helpful to determine the role of van der Waals and Casimir forces in
stability of thin films, which is the problem important for numerous applications \cite{55}.

We calculate the Casimir free energy and pressure for Au and Ag films deposited on the
plates made of different metals. All calculations are done using the Drude and plasma
models for the description of conduction (free) electrons. The contribution of core
(bound) electrons is taken into account by means of the tabulated optical data.
It is shown that the Casimir free energy of a film can be negative, positive or change
its sign as a function of film thickness depending on the film and plate metals.
We demonstrate that even for thin films of several tens of nanometers thickness
the Casimir free energy and pressure computed using the Drude and plasma model
approaches can differ by hundreds and thousands percent. This result is important
for resolving the Casimir puzzle formulated above.

One more qualitative result is that the Casimir free energy of metal film on a metallic
plate reaches the classical limit for rather small film thickness of about 150\,nm and
only if the Drude model is used for the theoretical description of free electrons.
If the plasma model is used for this purpose, the free energy drops exponentially fast
to zero with increasing film thickness, and there is no classical limit. Because of this,
starting from rather small film thicknesses, the use of optical data leads to only
minor influence on the computational results obtained by using the simple Drude model.
For thin films, however, the optical data contribute essentially, especially if the
Casimir energy of a film changes its sign. We show that in the framework of the
plasma model approach the optical data should be taken into account for films of any
thickness. It is shown also that the free energy of a metal film does not vanish in the
ideal metal limit if the Drude model is used.

The paper is organized as follows. Section~II presents general formalism of the Lifshitz
theory for metal films deposited on metallic plates and some analytic results.
In Sec.~III the computational results using the tabulated optical data and the Drude
and plasma models are presented for the film and plate metals leading to the negative
free energies and pressures. In Sec.~IV similar results are given
for the film and plate metals leading to the positive free energies and pressures.
Section~V is devoted to cases where the Casimir free energy changes its sign.
Section~VI contains our conclusions and discussion.

\section{General formalism for two-layer metallic system}

The case of a two-layer system is most simply obtained from the three-layer case
when one of the dielectric permittivities is put equal to unity.
Let the dielectric permittivities of a thick metallic plate (semispace) and
deposited on it metal film of thickness $a$ be $\varepsilon^{(1)}(\omega)$ and
$\varepsilon^{(2)}(\omega)$, respectively. The third (thick) layer we replace
with a vacuum and put $\varepsilon^{(3)}(\omega)=1$. Assuming that out two-layer
system is in thermal equilibrium with an environment at temperature $T$, for the
Casimir free energy of a metal film per unit area one obtains \cite{2,18,54}
\begin{equation}
{\cal F}(a,T)=\frac{k_BT}{2\pi}\sum_{l=0}^{\infty}{\vphantom{\sum}}^{\prime}
\int_{0}^{\infty}k_{\bot}\,dk_{\bot}
\sum_{\alpha}\ln\left[1-r_{\alpha}^{(2,3)}(i\xi_l,k_{\bot})
r_{\alpha}^{(2,1)}(i\xi_l,k_{\bot})e^{-2ak_l^{(2)}(k_{\bot})}\right].
\label{eq1}
\end{equation}
\noindent
Here, $k_B$ is the Boltzmann constant,
$\xi_l=2\pi k_BTl/\hbar$ with $l=0,\,1,\,2,\,\ldots$ are the
Matsubara frequencies,
$k_{\bot}=|\mbox{\boldmath$k$}_{\bot}|$ is the magnitude of the
projection of the wave vector on the plane of plates,
the prime adds a multiple 1/2 to the term with $l=0$, and the second
summation is made over two independent
polarizations of the electromagnetic field, transverse magnetic
($\alpha={\rm TM}$) and transverse electric ($\alpha={\rm TE}$).
The respective reflection coefficients  are given by
\begin{equation}
r_{\rm TM}^{(2,n)}(i\xi_l,k_{\bot})=\frac{\varepsilon_{l}^{(n)}
k_{l}^{(2)}(k_{\bot})-\varepsilon_{l}^{(2)}
k_l^{(n)}(k_{\bot})}{\varepsilon_{l}^{(n)}
k_{l}^{(2)}(k_{\bot})+\varepsilon_{l}^{(2)}
k_l^{(n)}(k_{\bot})},
\quad
r_{\rm TE}^{(2,n)}(i\xi_l,k_{\bot})=\frac{k_{l}^{(2)}(k_{\bot})-
k_l^{(n)}(k_{\bot})}{k_{l}^{(2)}(k_{\bot})+
k_l^{(n)}(k_{\bot})},
\label{eq2}
\end{equation}
\noindent
where $n =1,\,2,\,3$ and the following notation is introduced:
\begin{equation}
k_l^{(n)}(k_{\bot})=\sqrt{k_{\bot}^2+
{\varepsilon_{l}^{(n)}}\frac{\xi_l^2}{c^2}}
\label{eq3}
\end{equation}
\noindent
and $\varepsilon_{l}^{(n)}\equiv\varepsilon^{(n)}(i\xi_l)$.

{}From Eq.~(\ref{eq1}) for the Casimir pressure on a metallic film one obtains
\begin{equation}
P(a,T)=-\frac{k_BT}{\pi}\sum_{l=0}^{\infty}{\vphantom{\sum}}^{\prime}
\int_{0}^{\infty}k_{\bot}k_l^{(2)}(k_{\bot})\,dk_{\bot}
\sum_{\alpha}\left[
\frac{e^{2ak_l^{(2)}(k_{\bot})}}{r_{\alpha}^{(2,3)}(i\xi_l,k_{\bot})
r_{\alpha}^{(2,1)}(i\xi_l,k_{\bot})}
-1\right]^{-1}.
\label{eq4}
\end{equation}

Taking into account the present state of affairs (see Sec.~I), we describe the
dielectric permittivities of the film and plate metals by the Drude or
plasma model (the bound electrons are taken into account in numerical computations
of Secs.~III--V). The dielectric permittivity of the Drude model
at real and imaginary Matsubara
frequencies takes the form
\begin{equation}
\varepsilon_D^{(n)}(\omega)=1-\
\frac{\omega_{p,n}^2}{\omega(\omega+i\gamma_n)}, \quad
\varepsilon_{l,D}^{(n)}=1+\
\frac{\omega_{p,n}^2}{\xi_l(\xi_l+\gamma_n)},
\label{eq5}
\end{equation}
\noindent
where $\omega_{p,n}$ are the plasma frequencies of the film ($n=2$) and plate
($n=1$) metals and $\gamma_n$ are their relaxation frequencies (the latter are
temperature-dependent, and all computations below are performed at room temperature
$T=300\,$K). The dielectric permittivity of the plasma model is obtained from
Eq.~(\ref{eq5}) by putting $\gamma_n=0$
\begin{equation}
\varepsilon_p^{(n)}(\omega)=1-\
\frac{\omega_{p,n}^2}{\omega^2}, \quad
\varepsilon_{l,p}^{(n)}=1+\
\frac{\omega_{p,n}^2}{\xi_l^2},
\label{eq6}
\end{equation}
\noindent
Note that both dielectric functions (\ref{eq5}) and (\ref{eq6}) can be analytically
continued to the entire plane of complex frequencies and satisfy the Kramers-Kronig
relations formulated for the functions having the first- and second-order poles at
zero frequency, respectively \cite{56}.

We start from the zero-frequency contribution to Eqs.~(\ref{eq1}) and (\ref{eq4}).
In the case of the Drude model (\ref{eq5}), from Eq.~(\ref{eq2}) one finds
\begin{equation}
r_{{\rm TM},D}^{(2,3)}(0,k_{\bot})=-1, \quad
r_{{\rm TE},D}^{(2,n)}(0,k_{\bot})=0,
\quad
r_{{\rm TM},D}^{(2,1)}(0,k_{\bot})=\frac{\omega_{p,1}^2\gamma_2-
\omega_{p,2}^2\gamma_1}{\omega_{p,1}^2\gamma_2+\omega_{p,2}^2\gamma_1}
\equiv r_{D}^{(0)}.
\label{eq7}
\end{equation}
\noindent
As is seen in Eq.~(\ref{eq7}), these reflection coefficients do not depend on
$k_{\bot}$ and $|r_D^{(0)}|\ll 1$. Note that the quantity $r_D^{(0)}$ may be
both positive and negative depending on the relationship between the Drude
parameters of a film and a plate metals. Substituting  Eq.~(\ref{eq7}) in
Eq.~(\ref{eq1}), for the zero-frequency contribution to the Casimir free energy
calculated using the Drude model we have
\begin{equation}
{\cal F}_{D}^{(l=0)}(a,T)=\frac{k_BT}{4\pi}
\int_{0}^{\infty}k_{\bot}dk_{\bot}
\ln\left(1+r_{D}^{(0)}e^{-2ak_{\bot}}\right).
\label{eq8}
\end{equation}

Introducing the new dimensionless integration variable $y=2ak_{\bot}$,
one arrives at
\begin{equation}
{\cal F}_{D}^{(l=0)}(a,T)=\frac{k_BT}{16\pi a^2}
\int_{0}^{\infty}y dy
\ln\left(1+r_{D}^{(0)}e^{-y}\right)
=-\frac{k_BT}{16\pi a^2}{\rm Li}_3(-r_D^{(0)}),
\label{eq9}
\end{equation}
\noindent
where ${\rm Li}_n(z)$ is the polylogarithm function. Depending on the sign of
$r_D^{(0)}$, the zero-frequency
 contribution to the free energy is either
positive or negative, i.e., corresponds to either repulsive or attractive
contribution to the force. Using the smallness of $r_D^{(0)}$, we can
represent Eq.~(\ref{eq9}) in the form of a series
\begin{equation}
{\cal F}_{D}^{(l=0)}(a,T)=\frac{k_BT}{16\pi a^2}
r_{D}^{(0)}\left[1-\frac{1}{8}r_{D}^{(0)}+
\frac{1}{27}(r_{D}^{(0)})^2-\ldots\,\right].
\label{eq10}
\end{equation}
Substituting  Eq.~(\ref{eq7}) in Eq.~(\ref{eq4}) and repeating similar calculation,
for the zero-frequency contribution to the Casimir pressure we find
\begin{equation}
P_{D}^{(l=0)}(a,T)
=-\frac{k_BT}{8\pi a^3}{\rm Li}_3(-r_D^{(0)})
=r_{D}^{(0)}\frac{k_BT}{8\pi a^3}\left[1-\frac{1}{8}r_{D}^{(0)}+
\frac{1}{27}(r_{D}^{(0)})^2-\ldots\,\right].
\label{eq11}
\end{equation}

According to Ref.~\cite{54}, the contribution of Matsubara terms with $l\geq 1$
to the Casimir free energy of a metal film sandwiched between two dielectric
plates is negligibly small for surprisingly thin films if the film metal is
described by the Drude model. By repeating the same calculation in our case
of a metal film deposited on a metallic plate, we find that the contribution of
all Matsubara terms with $l\geq 1$ becomes negligibly small for film thickness
$a>150\,$nm. For these film thicknesses Eqs.~(\ref{eq9})--(\ref{eq11}) represent
the total Casimir free energy and pressure, i.e., the classical limit is already
achieved (recall that in the configuration of two metallic plates separated with
a dielectric film the classical limit is achieved only at separations
exceeding $6\,\mu$m \cite{2}).

Now we consider the Casimir free energy and pressure ${\cal F}_D$ and $P_D$ in the
limiting case $\omega_{p,2}\to\infty$. This means that the dielectric permittivity
of a metal film becomes infinitely large at all frequencies, i.e., its material is
the so-called ideal metal. {}From Eqs.~(\ref{eq3}) and (\ref{eq5}) we conclude  that
all $k_l^{(2)}(k_{\bot})$ with $l\geq 1$ go to infinity when  $\omega_{p,2}\to\infty$.
Thus, according to Eqs.~(\ref{eq1}) and (\ref{eq4}), only the zero-frequency Matsubara
terms contribute to the free energy and pressure in this limiting case. Using
Eq.~(\ref{eq7}) we find that $r_D^{(0)}\to -1$ when $\omega_{p,2}\to\infty$. Then, from
Eqs.~(\ref{eq9}) and (\ref{eq11}) one obtains
\begin{equation}
{\cal F}_D(a,T)=-\frac{k_BT}{16\pi a^2}\zeta(3), \quad
P_D(a,T)=-\frac{k_BT}{8\pi a^3}\zeta(3),
\label{eq12}
\end{equation}
\noindent
where $\zeta(z)$ is the Riemann zeta function. The results (\ref{eq12}) mean that an ideal
metal film is characterized by a nonzero Casimir free energy and pressure if the Drude model
is used in calculations. This is paradoxical if to take into account that electromagnetic
fluctuations cannot penetrate in the interior of ideal metal.

If metals of a film and a plate are described by the plasma model (\ref{eq6}), one arrives
at quite different results. Substituting Eq.~(\ref{eq6}) in the first line of Eq.~(\ref{eq2}),
the TM reflection coefficients at zero Matsubara frequency take the form
\begin{equation}
r_{{\rm TM},p}^{(2,1)}(0,k_{\bot})=\frac{(\omega_{p,1}^2-\omega_{p,2}^2)[c^2k_{\bot}^2
(\omega_{p,1}^2+\omega_{p,2}^2)+\omega_{p,1}^2\omega_{p,2}^2]}{(\omega_{p,1}^2
\sqrt{c^2k_{\bot}^2+\omega_{p,2}^2}+\omega_{p,2}^2\sqrt{c^2k_{\bot}^2+\omega_{p,1}^2})^2},
\quad
r_{{\rm TM},p}^{(2,3)}(0,k_{\bot})=-1.
\label{eq13}
\end{equation}
\noindent
It can be seen that $r_{{\rm TM},p}^{(2,1)}$ is positive for $\omega_{p,1}>\omega_{p,2}$ and
negative  for $\omega_{p,1}<\omega_{p,2}$.
{}From the second line of Eq.~(\ref{eq2}) we have
\begin{eqnarray}
&&
r_{{\rm TE},p}^{(2,1)}(0,k_{\bot})=
\frac{\omega_{p,2}^2-\omega_{p,1}^2}{(\sqrt{c^2k_{\bot}^2+\omega_{p,2}^2}+
\sqrt{c^2k_{\bot}^2+\omega_{p,1}^2})^2},
\nonumber \\
&&
r_{{\rm TE},p}^{(2,3)}(0,k_{\bot})=\frac{\omega_{p,2}^2}{(\sqrt{c^2k_{\bot}^2+\omega_{p,2}^2}+
ck_{\bot})^2}.
\label{eq14}
\end{eqnarray}
\noindent
As is seen in Eq.~(\ref{eq14}), $r_{{\rm TE},p}^{(2,1)}>0$ when $\omega_{p,2}>\omega_{p,1}$ and
$r_{{\rm TE},p}^{(2,1)}<0$ when $\omega_{p,2}<\omega_{p,1}$, whereas the reflection coefficient
$r_{{\rm TE},p}^{(2,3)}$ is always positive. The above results open possibilities for both
positive and negative Casimir free energy.

By repeating calculations of Ref.~\cite{54}, it can be proven that the Casimir free energy (\ref{eq1})
and pressure (\ref{eq4}) with increasing film thickness $a$ are expressed via the exponentially
small terms depending on the Planck constant $\hbar$. Thus, there is no classical limit in the
configuration of a metal film deposited on a metallic plate if the plasma model is used in
calculations.

In the limiting case of an ideal metal film ($\omega_{p,2}\to\infty$) from   Eq.~(\ref{eq3})
one finds
\begin{equation}
k_l^{(2)}(k_{\bot})=\sqrt{k_{\bot}^2+\frac{\xi_l^2}{c^2}+\frac{\omega_{p,2}^2}{c^2}}
\to\infty
\label{eq15}
\end{equation}
\noindent
for all $l\geq 0$.
Then, from Eqs.~(\ref{eq1}) and (\ref{eq4}) we obtain
\begin{equation}
{\cal F}(a,T)\to 0, \quad
P(a,T)\to 0,
\label{eq16}
\end{equation}
\noindent
as it should be in accordance to physical intuition. Below we consider all the above
general properties for several specific metals of a film and a plate.

\section{Films with negative Casimir free energy}

Here, we compute the Casimir free energy and pressure for Ag and Au films deposited
on Cu plates using Eqs.~(\ref{eq1})--(\ref{eq4}).
The dielectric permittivities of these metals at the imaginary Matsubara frequencies
are obtained by means of the Kramers-Kronig relations using the tabulated optical data
\cite{57} extrapolated to zero frequency either by the Drude of by the plasma models
(the Drude and plasma model approaches, respectively). In the case of Au such
extrapolations with the plasma frequency
$\omega_{p,{\rm Au}}=9.0\,$eV and relaxation parameter  $\gamma_{\rm Au}=0.035\,$eV
have been considered in detail and extensively used in calculations of the Casimir
force and comparison with the experimental data \cite{2,16}.
For Ag the optical data \cite{57} extend over the same frequency interval as for Au,
i.e., from 0.125 to $10^4\,$eV, and the Drude parameters of the extrapolation are
$\omega_{p,{\rm Ag}}=9.66\,$eV and $\gamma_{\rm Ag}=0.0315\,$eV
in agreement with Ref.~\cite{58}. For Cu the optical data \cite{57} extend
from 0.13 to $9\times10^3\,$eV and similar extrapolation to lower frequencies results in
$\omega_{p,{\rm Cu}}=8.6\,$eV and $\gamma_{\rm Cu}=0.0325\,$eV.
These results are rather close to the Drude parameter of Cu in Refs.~\cite{59,60}.
Taking into account that for all three metals the tabulated optical data extend up to
sufficiently high frequencies, there is no need in further interpolations.

\begin{figure}[b]
\vspace*{-2cm}
\centerline{\hspace*{0cm}
\includegraphics{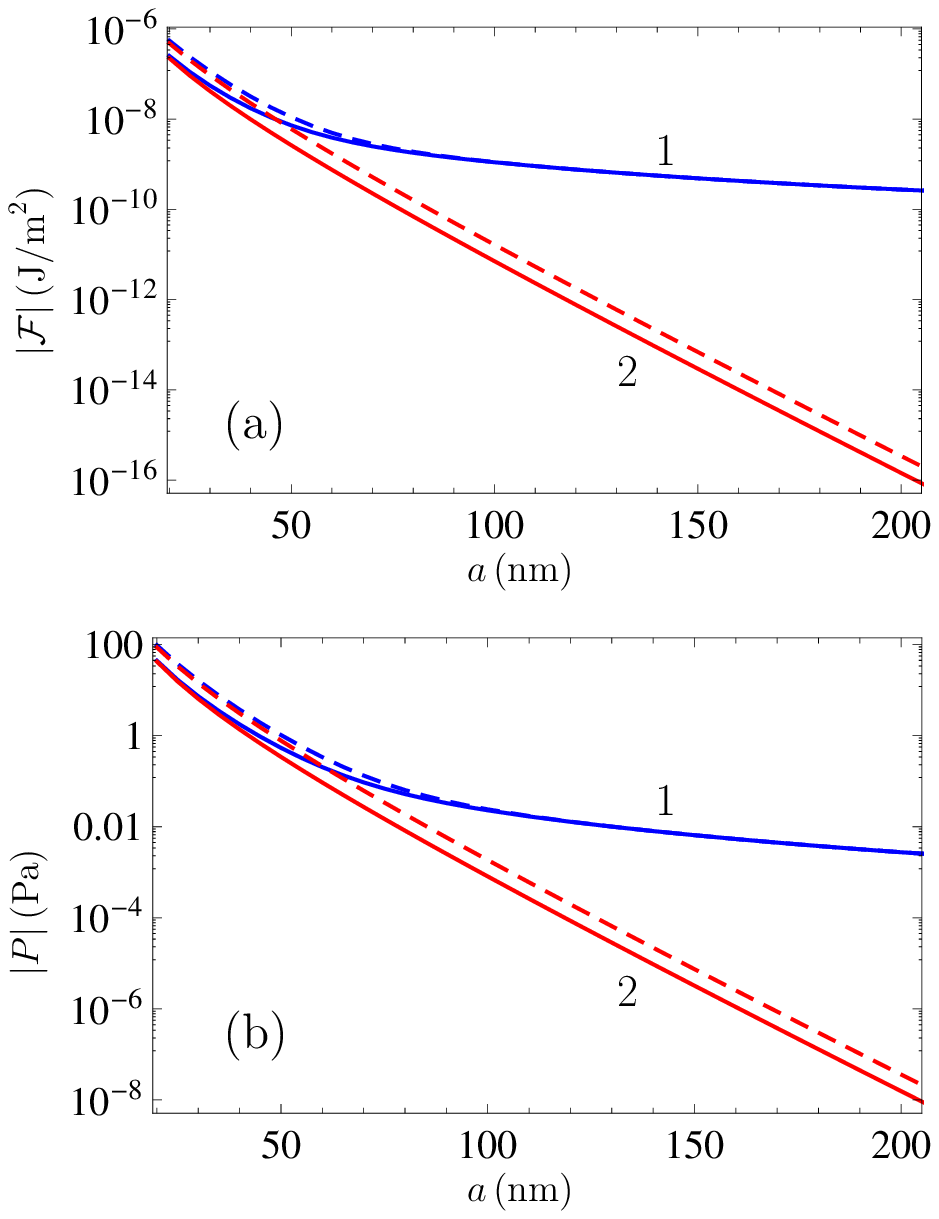}
}
\vspace*{-13cm}
\caption{\label{fg1}(Color online)
The  magnitudes of (a) the Casimir free energy per unit area and (b) the Casimir pressure of
a Ag film on a Cu plate computed using the tabulated optical data extrapolated to zero
frequency by the Drude (the solid line 1) and plasma (the solid line 2) models are shown
as functions of the film thickness. The dashed lines 1 and 2 present the same quantities
computed using the simple Drude and plasma models, respectively.
}
\end{figure}

In Fig.~\ref{fg1}(a) we present the computational results for the magnitude of the
Casimir free energy per unit area of a Ag film deposited on a Cu plate as a function
of film thickness $a$ at $T=300\,$K (all subsequent calculations are also performed at
room temperature). Computations are performed over the range of $a$ from 20 to 200\,nm.
Note that for $a>10\,$nm one can neglect by the effect of anisotropy of atomically
thin planar layers \cite{61,62}. The solid and dashed lines labeled 1 are computed
using the optical data of both Ag and Cu extrapolated by the Drude model and the
simple Drude model (\ref{eq5}), respectively. The solid and dashed lines labeled 2
are calculated
using the optical data of both Ag and Cu extrapolated by the plasma model and the
simple plasma model (\ref{eq6}), respectively.

According to our computational results, the Casimir free energy of Ag films on a Cu plate is
always negative, i.e., the respective forces are attractive. Below we demonstrate that this
result is not universal, so that the sign of ${\cal F}$ depends on the metals of a film and
a plate and on the film thickness. As can be seen in Fig.~\ref{fg1}(a), the magnitudes
of the Casimir free energy calculated using the optical data extrapolated by the Drude
and plasma models are significantly different. Thus, for Ag films of 50 and 100\,nm
 thickness the Casimir free energies predicted by the Drude and plasma model
approaches differ by the factors of 2.76 and 156.6, respectively. With increasing film
thickness $a$ the difference in theoretical predictions of both approaches further
increases. This is explained by the exponentially fast vanishing of the Casimir free
energy calculated using the plasma model approach (see the solid line 2), whereas
${\cal F}_D$ (the solid line 1) goes to the classical limit (\ref{eq9}). We remind that
in the configuration of two metallic plates separated with a vacuum gap of width $a$
the difference in theoretical predictions of the Drude and plasma model approaches
at $a<1\,\mu$m is below a few percent and reaches 100\% only at $a=6\,\mu$m  \cite{2,16}.

We also note on the role of optical data in both calculation approaches.
As is seen in Fig.~\ref{fg1}(a), an influence of the optical data on computational results
is much larger in the plasma model approach than in the Drude model one. Thus, for the films
of 50 and 100\,nm thickness the simple Drude model (the dashed line 1) leads to by the factors
1.52 and 1.01 larger results, respectively, than the optical data extrapolated by the Drude
model (the solid line 1). For the plasma model approach the respective factors found from a
comparison of the dashed and solid lines 2 are 2.30 and 2.31, and they do not decrease with
increasing $a$. This is explained by the fact that for the plasma model approach the
relative role of the zero-frequency term (which does not depend on the optical data)
decreases with increasing film thickness. For two metallic plates separated with a vacuum gap
the Casimir free energy is always negative, and the magnitudes of the Casimir free energy
computed using the optical data are larger than using the simple Drude or plasma models.
In that case the optical data do not influence as strong as for a metal film on a metallic
plate. Thus, for a Ag plate interacting with a Cu plate at separations 50 and 100\,nm
an excess of $|{\cal F}|$ due to the use of optical data is by the factors of 1.094 and 1.037
for the Drude model and of 1.097  and 1.038 for the plasma model, respectively.

In Fig.~\ref{fg1}(b) the solid and dashed lines labeled 1 present the Casimir pressure
in a Ag film computed
using the optical data  extrapolated by the Drude model and the
simple Drude model, respectively, as a function
of the film thickness. The solid and dashed lines labeled 2 show similar results
computed using the plasma model. As is seen in Fig.~\ref{fg1}(b), the Casimir pressure
behaves analogously to the free energy. Thus, the role of optical data is more pronounced
in the plasma model approach than in the Drude model one. With increasing $a$ the Casimir
pressure computed using the plasma model exponentially fast drops to zero, i.e., there is
no classical limit. For the Drude model approach the classical limit (\ref{eq11}) is
already achieved for films of 150\,nm thickness. Similar to the case of the free energy,
the Casimir pressures calculated using the Drude and plasma model approaches are
significantly different even for Ag films of 50 and 100\,nm thickness.

\begin{figure}[b]
\vspace*{-7cm}
\centerline{\hspace*{1cm}
\includegraphics{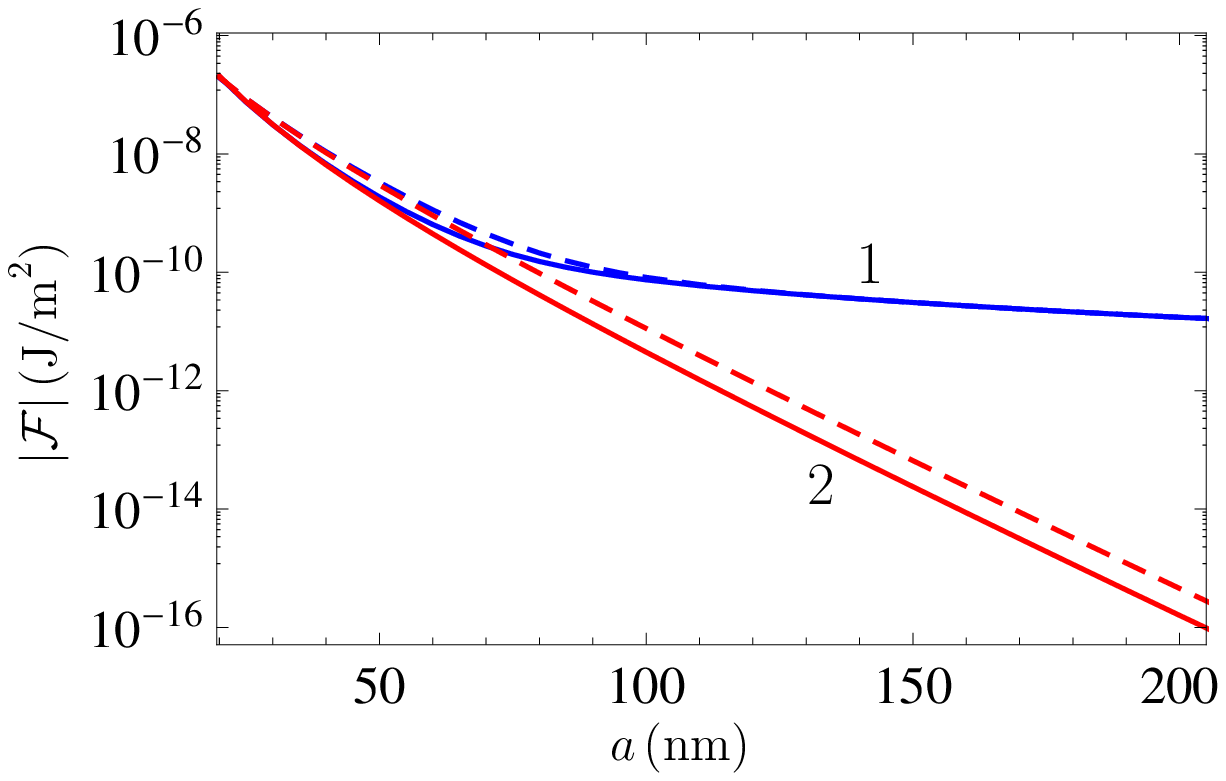}
}
\vspace*{-9cm}
\caption{\label{fg2}(Color online)
The  magnitudes of the Casimir free energy per unit area  of
a Au film on a Cu plate computed using the tabulated optical data extrapolated to zero
frequency by the Drude (the solid line 1) and plasma (the solid line 2) models are shown
as functions of the film thickness. The dashed lines 1 and 2 present the same quantities
computed using the simple Drude and plasma models, respectively.}
\end{figure}

For a Ag film deposited on a Cu plate the respective plasma frequencies satisfy an inequality
$\omega_{p,{\rm Ag}}>\omega_{p,{\rm Cu}}$. One more example of this kind is presented by a Au
film deposited on a Cu substrate. The magnitude of the Casimir free energy of this film
computed using the tabulated optical data extrapolated by the Drude and plasma models
are shown in Fig.~\ref{fg2} as functions of film thickness by the solid lines labeled 1 and 2,
respectively. The dashed lines 1 and 2 are computed using the simple Drude and plasma models,
respectively. Similar to the case of a Ag film  on a Cu plate, the free energy of a Au film
is negative. The classical limit is reached for Au films of about 150\,nm thickness
if the Drude model approach is used. There is no classical limit in the plasma model approach.
As is seen in Fig.~\ref{fg2},  for films of 50 and 100\,nm thickness the magnitudes of the
free energy computed using the optical data and the Drude model are by the factors of 1.15 and
16.6 larger than those computed using the optical data and the plasma model, respectively.
These factors are smaller than for a Ag film but still much larger than for two metallic plates
interacting through a vacuum gap. The role of optical data in the Drude model approach decreases
with increasing film thickness. For instance, for films of 50 and 100\,nm thickness the
magnitudes of the free energy computed using the simple Drude model are larger by the factors
of 1.76 and 1.10 than those computed using the optical data, respectively.
For the plasma model approach the respective factors are 1.84 and 2.50 for the same film
thicknesses. Here, the role of optical data does not decrease with increasing film thickness.

\section{Films with positive Casimir free energy}

Unlike the case of two metallic plates interacting through a vacuum gap, the free energy and
pressure of a metal film deposited on a metallic plate can be positive. Here, we consider
Au and Ag films on an Al plate. The tabulated optical data for Al over the frequency range
from 0.04 to $10^4\,$eV were taken in Ref.~\cite{57} and extrapolated to zero frequency by the
Drude or plasma models with the parameters $\omega_{p,{\rm Al}}=11.34\,$eV and
$\gamma_{\rm Al}=0.041\,$eV in rather good agreement with Ref.~\cite{63}.
Then the dielectric permittivity of Al at the imaginary Matsubara frequencies using the
Drude and plasma model approaches was found as described in Refs.~\cite{2,16}. For the films
considered in this section the plasma frequencies are less than the plasma frequency of the
plate. This makes possible the effect of the Casimir repulsion.

\begin{figure}[b]
\vspace*{-2cm}
\centerline{\hspace*{0cm}
\includegraphics{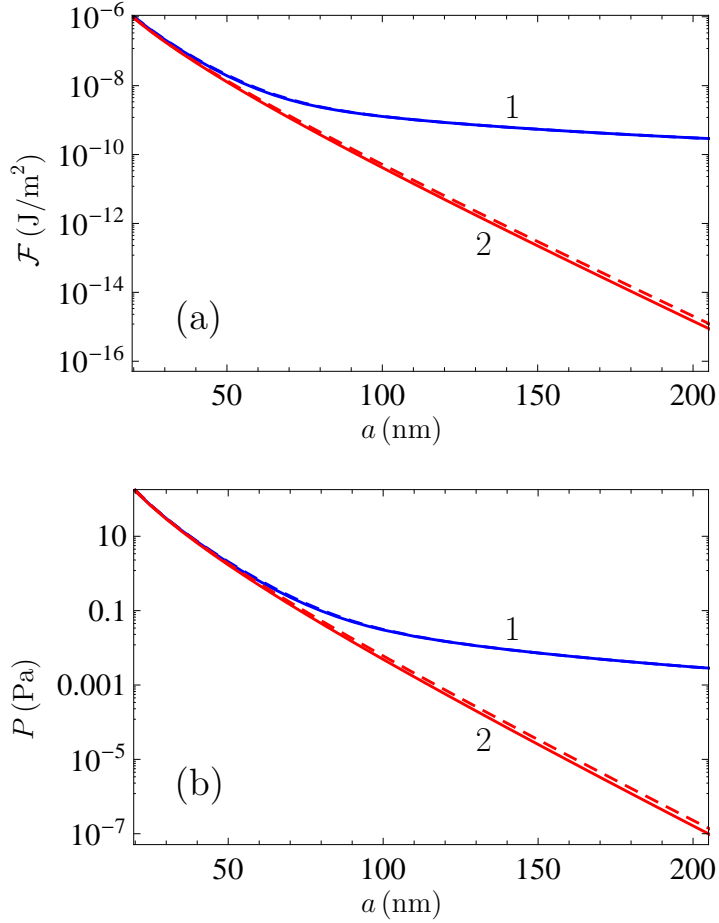}
}
\vspace*{-13cm}
\caption{\label{fg3}(Color online)
(a) The Casimir free energies per unit area and (b) the Casimir pressures of
a Au film on an Al plate computed using the tabulated optical data extrapolated to zero
frequency by the Drude (the solid line 1) and plasma (the solid line 2) models are shown
as functions of the film thickness. The dashed lines 1 and 2 present the same quantities
computed using the simple Drude and plasma models, respectively.}
\end{figure}

In Fig.~\ref{fg3}(a) the solid and dashed lines labeled 1 present the Casimir free energy
per unit area of a Au film on an Al plate computed as a function of film thickness using
the optical data and the Drude model (the solid line) and the simple Drude model (the dashed
line). Similar results computed using the optical data and the plasma model or the simple
plasma model  are shown by the solid and dashed lines labeled 2, respectively.
In Fig.~\ref{fg3}(b) the computational results for the Casimir pressure
of a Au film on an Al plate are presented using the same notations.
As is seen in Fig.~\ref{fg3}(a,b), the Casimir free energy and pressure of a Au film in this
case are positive which corresponds to the effect of repulsion. In the same manner as in
Figs.~\ref{fg1} and \ref{fg2}, the Drude and plasma model approaches predict significantly
different Casimir free energies (the solid lines 1 and 2). For film thicknesses of 50 and
100\,nm these predictions differ by the factors of 1.52 and 30.6, respectively. With further
increase of the film thickness the difference in theoretical predictions reaches several
orders of magnitude. This is again explained by the exponentially fast decreasing of the
Casimir free energy and pressure when the plasma model approach is used, whereas in the
framework of the Drude model approach both these quantities reach the classical limit for
film thicknesses of about 150\,nm.
The role of optical data in Fig.~\ref{fg3} is evidently smaller than in  Figs.~\ref{fg1}
and \ref{fg2}. For the dashed and solid lines labeled 1 (computed using the simple Drude
model and the optical data extrapolated by the Drude model) the ratio of respective free
energies for $a=50\,$nm is equal to 1.057 and unity for $a=100\,$nm.
For the dashed and solid lines labeled 2 (computed using the simple plasma
model and the optical data extrapolated by the plasma model) the respective ratios
take the values 1.094 and 1.234. Similar results are obtained for the Casimir pressure.

\begin{figure}[b]
\vspace*{-7cm}
\centerline{\hspace*{1cm}
\includegraphics{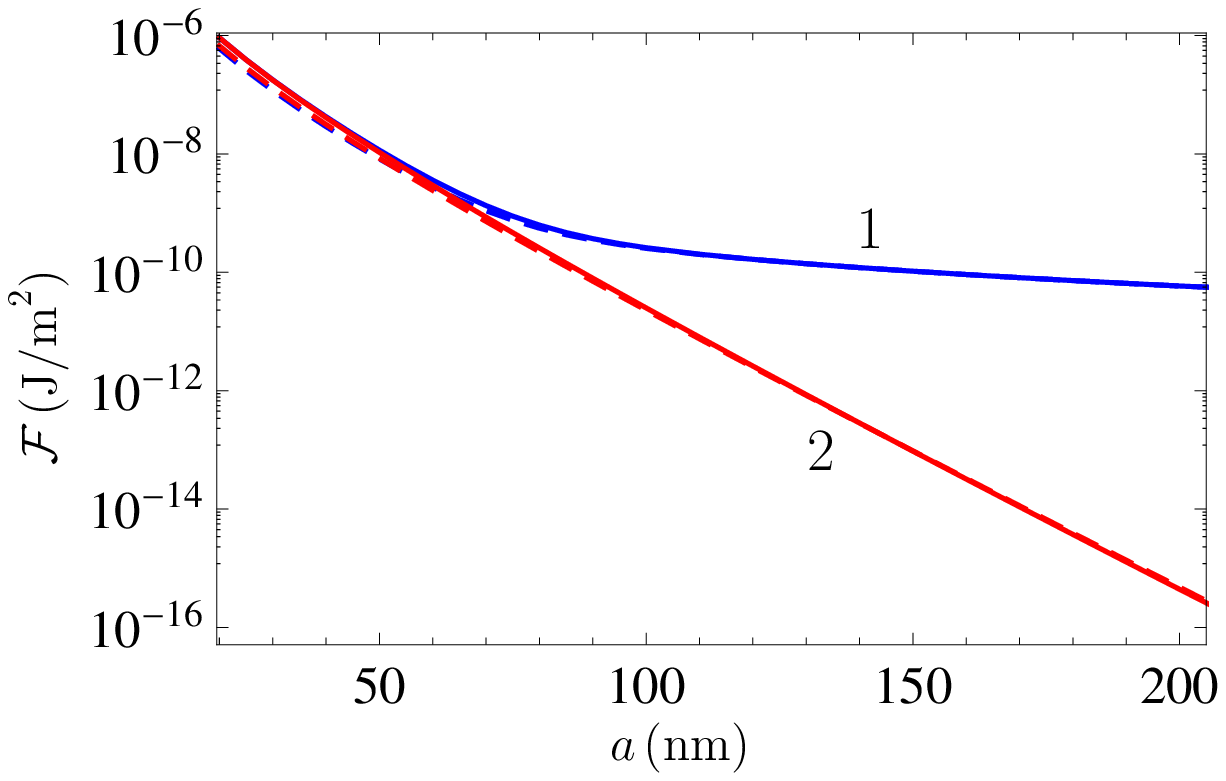}
}
\vspace*{-9cm}
\caption{\label{fg4}(Color online)
The   Casimir free energies per unit area of
a Ag film on an Al plate computed using the tabulated optical data extrapolated to zero
frequency by the Drude (the solid line 1) and plasma (the solid line 2) models are shown
as functions of the film thickness. The dashed lines 1 and 2 present the same quantities
computed using the simple Drude and plasma models, respectively.}
\end{figure}

One more example of the positive Casimir free energy is given by a Ag film deposited on
an Al plate. In this case the Casimir free energy calculated using the Drude and plasma
model approaches is presented in Fig.~\ref{fg4} as a function of film thickness by the
solid lines 1 and 2, respectively. The dashed lines 1 and 2 show the Casimir free energy
calculated using the simple Drude and plasma models. Similar to Fig.~\ref{fg3}, here the
Casimir free energy is positive. The distinctions of Fig.~\ref{fg4} from  Fig.~\ref{fg3}(a)
are, however, not of only quantitative character. Thus, for a Ag film on an Al plate the free
energies computed using the optical data are larger than the ones computed using the
simple Drude and plasma models. For Ag films of 50 and 100\,nm thickness the free
energies computed using the optical data are larger by the factors of 1.396 and 1.024 if
the Drude model is used and by the factors of 1.259 and 1.092 in the case of the plasma
model. The Casimir free energies of Ag films of 50 and 100\,nm thickness computed using
the Drude model approach (the solid line 1) are larger than those computed using
the plasma model approach (the solid line 2) by the factors of 1.085 and 10.4, respectively.

\section{Films with varying sign of the Casimir free energy}

\begin{figure}[b]
\vspace*{-2cm}
\centerline{\hspace*{0cm}
\includegraphics{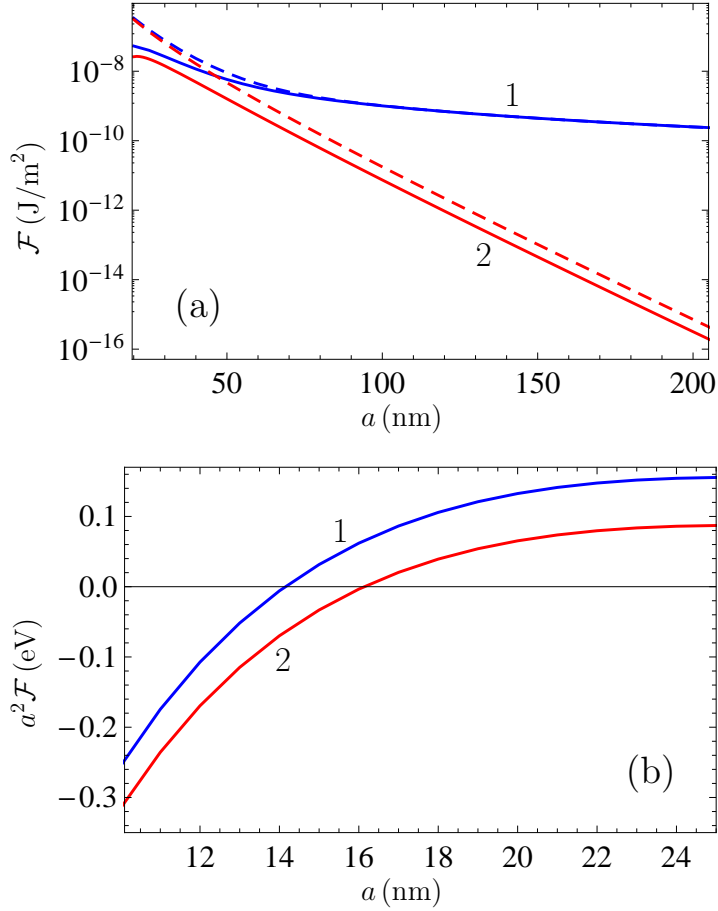}
}
\vspace*{-13cm}
\caption{\label{fg5}(Color online)
(a) The   Casimir free energies per unit area of a Au film on a Ag plate
and (b) the same free energies multiplied by the film thickness squared
computed using the tabulated optical data extrapolated to zero
frequency by the Drude (the solid line 1) and plasma (the solid line 2) models are shown
as functions of the film thickness. The dashed lines 1 and 2 present the free energies
computed using the simple Drude and plasma models, respectively.}
\end{figure}

In this section we demonstrate that for some pairs of metals the Casimir free energy of a film
changes its sign depending on the film thickness. We start from a Au film deposited on a Ag
plate. In this case the computational results for the positive Casimir free energy are
presented in Fig.~\ref{fg5}(a) by the pairs of lines labeled 1 and 2, where all the notations
are the same as in Figs.~\ref{fg1}--\ref{fg4}. For the relatively large film thickness the
free energy computed using the Drude model (the solid and dashed lines 1) goes to the
classical limit and computed using the plasma model (the solid and dashed lines 2)
exponentially fast drops to zero. For Au films of 50 and 100\,nm thickness the free energy
shown by the solid line 1 is by the factors of 36.25 and 135.9 larger than that shown by the
solid line 2, respectively. Similar to Figs.~\ref{fg1}--\ref{fg4} the role of the optical
data is more pronounced for the free energy computed using the plasma model approach.

For the smallest film thicknesses Fig.~\ref{fg5}(a), however, differs from all the above
figures. Here, the dashed lines 1 and 2 computed using the simple Drude and plasma models
deviate considerably from the solid lines 1 and 2 computed using the optical data.
The latter lines approach to their maximum values. Because of this, it is interesting to
calculate the Casimir free energy for thinner Au films in the range from 10 to 20\,nm,
where the role of optical data becomes dominant. In Fig.~\ref{fg5}(b) the computational
results for ${\cal F}$ multiplied by the second power of the film thickness are shown by the
solid lines 1 and 2 computed using the optical data extrapolated by the Drude and plasma
models, respectively.  As is seen in Figs.~\ref{fg5}(a,b), the maximum values of ${\cal F}$
are reached for films of 19 and 21\,nm thickness when the Drude and plasma model approaches
are used in computations, respectively. {}From Fig.~\ref{fg5}(b) it is seen that with
decreasing film thickness the
Casimir free energy decreases and takes the zero value for $a\approx 14.2$ and $16.1\,$nm
for the Drude and plasma model approaches, respectively.  For thinner films the Casimir free energy
changes its sign from plus to minus, i.e., the respective Casimir pressure becomes attractive
from repulsive. If for some film thickness the free energy predicted by one of the calculation
approaches is almost zero, a discrepancy between the predicted values increases. Thus, for
Au films of 14 and 16\,nm  thickness the magnitudes of the free energy calculated using
the Drude and plasma model approaches differ by the factors of 11.8 and 18.0, respectively.

\begin{figure}[b]
\vspace*{-2cm}
\centerline{\hspace*{0cm}
\includegraphics{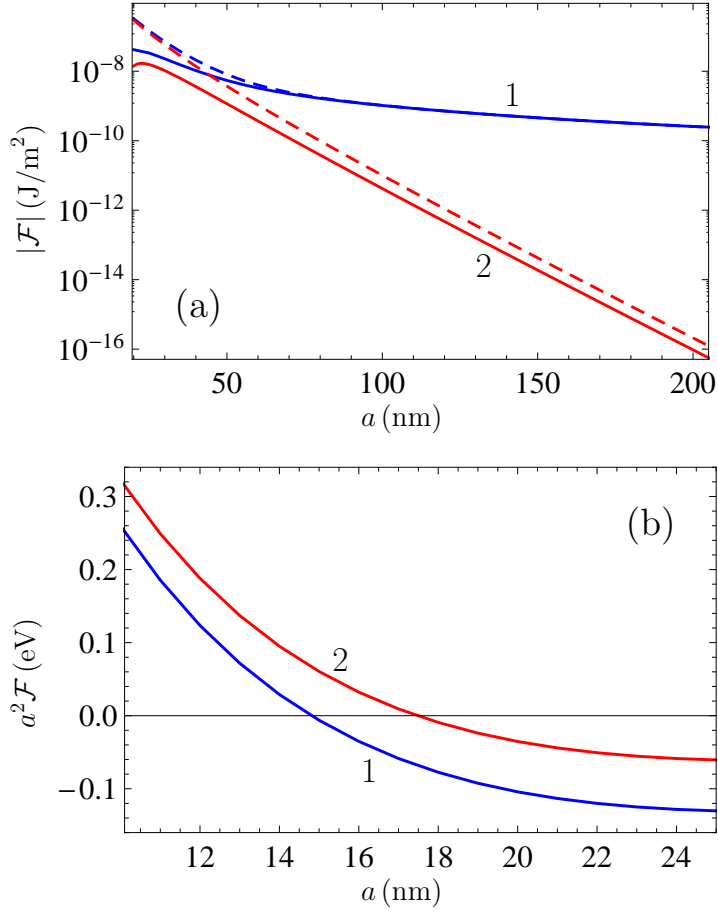}
}
\vspace*{-13cm}
\caption{\label{fg6}(Color online)
(a) The  magnitudes of the Casimir free energy per unit area of a Ag film on a Au plate
and (b) the Casimir free energies multiplied by the film thickness squared
computed using the tabulated optical data extrapolated to zero
frequency by the Drude (the solid line 1) and plasma (the solid line 2) models are shown
as functions of the film thickness. The dashed lines 1 and 2 present the
magnitudes of the free energies
computed using the simple Drude and plasma models, respectively.}
\end{figure}

It is also possible to get the positive free energy (i.e., the Casimir repulsion) for very
thin films and change its sign for the negative one with increasing film thickness. For this
purpose one can use Ag films deposited on a Au plate. In Fig.~\ref{fg6}(a) the magnitude of
the negative Casimir free energy  of a Ag film on a Au plate is plotted as a function of
the film thickness using the same notations as in all previous figures.
For the films of 50 and 100\,nm thickness the magnitude of the free energy computed by the
Drude model approach (the solid line 1) is larger than that computed by the
plasma model approach (the solid line 2)
by the factors of 46.8 and 245.9, respectively.
Again, with decreasing film thickness the dashed lines 1 and 2 computed using the simple
Drude and plasma models deviate significantly from the solid lines 1 and 2 which approach to
their maximum values reached for $a=20$ and 23\,nm, respectively.

In Fig.~\ref{fg6}(b) we plot the Casimir free energy of a Ag film on a Au plate computed using the
Drude and plasma model approaches for film thicknesses in the range from 10 to 25\,nm.
In order to use the natural scale, the quantity  $a^2{\cal F}$ is plotted rather than ${\cal F}$.
As is seen in Fig.~\ref{fg6}(b), the Casimir free energies computed using
the Drude and plasma model approaches
take the zero value for films of approximately $14.8$ and $17.5\,$nm thickness, respectively.
For thinner Ag films the Casimir free energy changes its sign and becomes positive, which
correspond to the Casimir repulsion.
For Ag films of 15 and 18\,nm  thickness the magnitudes of the Casimir free energy predicted by
the Drude and plasma model approaches are different by the factors of 9.6 and 8.4, respectively.

\section{Conclusions and discussion}

In the foregoing we have investigated the Casimir free energies and pressures of metal films deposited
on metallic plates made of different metals. It is shown that this configuration possesses unusual
properties which are significantly different from the well studied case of two metallic plates
interacting through a vacuum gap. These properties shed new light on an unresolved problem in
Casimir physics, the so-called Casimir puzzle, and are also interesting for applications because
the van der Waals and Casimir forces play an important role in the stability of thin films.

Specifically, we have shown that the Casimir free energy of metal films of a few tens nanometer
thickness deposited on a metallic plate  differs widely depending on whether the Drude or the
plasma model approach is employed in calculations. Depending on metals used this difference can reach
hundreds and even thousands of percent, so that it can be easily discriminated experimentally.
One more qualitative result is that the free energy of a metal film on a metallic plate reaches
the classical limit with increasing film thickness only if the Drude model approach is used
in computations. This limit is achieved for unusually small film thickness of about 150\,nm.
As a result, in the limiting case of an ideal metal the Casimir free energy of a metal film
does not vanish in contradiction with the fact that electromagnetic fluctuations cannot
penetrate in its interior. If the film and plate metals are described by the plasma model approach,
the Casimir free energy of a film drops to zero exponentially fast and there is no classical limit.

We have performed numerical computations of the Casimir free energy and pressure of Ag and Au films
deposited on Cu and Al plates using the Drude and plasma model approaches. It is shown that for
a Cu plate  the Casimir free energy and pressure of Ag and Au films are negative and correspond
to the Casimir attraction, whereas for an Al plate they are positive and correspond
to the Casimir repulsion. For a Au film deposited on a Ag plate and for a Ag film deposited on
a Au plate the Casimir free energy changes its sign from minus to plus and from plus to minus,
respectively, with increasing film thickness. All computations have been made in two different
ways: using the complete optical data for the complex index of refraction of the film and plate
metals extrapolated to zero frequency by  the Drude and plasma models and by using the simple
Drude and plasma models at all frequencies. According to our results, in the framework of the
Drude model approach the optical data influence the Casimir free energy and pressure of thin
metal films, and this influence quickly decreases for films from 50 to 100\,nm thickness
depending on the film and plate metals. If the plasma model approach is used in computations,
the optical data affect considerably the obtained values of the free energy and pressure for
any film thickness.

To conclude, the configuration of metal films deposited
on metallic plates display several unusual properties of
the Casimir free energy and pressure which makes it interesting for various applications
in both fundamental physics and in surface science.


\newpage
\end{document}